\documentclass[sigconf,anonymous=false,review=false]{acmart}

\copyrightyear{2021}
\acmYear{2021}
\setcopyright{iw3c2w3}
\acmConference[WWW '21 Companion]{Companion Proceedings of the Web Conference 2021}{April 19--23, 2021}{Ljubljana, Slovenia}
\acmBooktitle{Companion Proceedings of the Web Conference 2021 (WWW '21 Companion), April 19--23, 2021, Ljubljana, Slovenia}
\acmPrice{}
\acmDOI{10.1145/3442442.3451887}
\acmISBN{978-1-4503-8313-4/21/04}

\usepackage{url}  %
\usepackage{graphicx}  %
\usepackage{booktabs} %

\usepackage{hyperref}
\usepackage{tikz}
\usepackage{pgfplots}
\pgfplotsset{compat=1.13}
\usepgfplotslibrary{colormaps}
\usetikzlibrary{pgfplots.colormaps}
\usetikzlibrary{pgfplots.dateplot}
\usetikzlibrary{patterns}
\usepackage{graphicx}
\usepackage{subcaption}
\usepackage{color, colortbl}
\usepackage{arydshln}
\usepackage{float}
\usepackage[utf8]{inputenc}
\usepackage{arabtex}
\usepackage{utf8}
\usepackage{units}
\usepackage{enumitem}
\usepackage{balance}
\usepackage{microtype} %
\usepackage[export]{adjustbox}
\usepackage{multirow}
\usepackage{tabularx}
\usepackage{hhline}
\usepackage{array,booktabs,multirow}
\newcolumntype{M}[1]{>{\centering\arraybackslash}m{#1}}
\usepackage{mwe}
\usepackage[export]{adjustbox}

\setcode{utf8}

\definecolor{Gray}{gray}{0.9}

\newcommand{\afblock}[1]{\noindent{\textbf{#1 }}}
\newcommand{\takeaway}[1]{\noindent{\textbf{Findings.}} \textit{#1}}

\newcommand\encircle[1]{%
	\tikz[
		baseline={([yshift=-8pt]current bounding box.north)}
	]
		\node (X) [draw, shape=circle, inner sep=0, fill=black, text=white,scale=0.7] {\strut #1};
	\hspace{-4pt}
}

\newcommand{\ie}{i.e.,\,}
\newcommand{\Ie}{I.e.,\,}
\newcommand{\eg}{e.g.,\,}
\newcommand{\Eg}{E.g.,\,}
\newcommand{\etal}{et~al\@ifnextchar.{}{.\@}}
\newcommand{\etc}{etc\@ifnextchar.{}{.\@}}

\newcommand{\wrt}{w.r.t.\@}
\newcommand{\sref}[1]{\S~\ref{#1}}


\begin{document}
\title[Predicting User Lifetime with ML in the Jodel Social Network]{Understanding \& Predicting User Lifetime with Machine Learning in an Anonymous Location-Based Social Network}

\author{Jens Helge Reelfs}
\affiliation{
  \institution{Chair for Computer Networks}
  \streetaddress{Konrad-Wachsmann-Allee 5}
  \city{Brandenburg University of Technology}
  \country{Germany}
  \postcode{03046}
}
\email{reelfs@b-tu.de}

\author{Max Bergmann}
\affiliation{
    \institution{Chair for Computer Networks}
    \streetaddress{Konrad-Wachsmann-Allee 5}
    \city{Brandenburg University of Technology}
    \country{Germany}
    \postcode{03046}
}
\email{max.bergmann@b-tu.de}

\author{Oliver Hohlfeld}
\affiliation{
  \institution{Chair for Computer Networks}
  \streetaddress{Konrad-Wachsmann-Allee 5}
  \city{Brandenburg University of Technology}
  \country{Germany}
  \postcode{03046}
}
\email{hohlfeld@b-tu.de}

\author{Niklas Henckell}
\affiliation{
  \institution{The Jodel Venture GmbH}
  \streetaddress{K\"openicker Str. 126}
  \city{Berlin}
  \country{Germany}
  \postcode{10179}
}
\email{niklas@jodel.com}

\begin{abstract}
In this work, we predict the user lifetime within the anonymous and location-based social network Jodel in the Kingdom of Saudi Arabia.
Jodel's location-based nature yields to the establishment of disjoint communities country-wide and enables for the first time the study of user lifetime in the case of a large set of disjoint communities.
A user's lifetime is an important measurement for evaluating and steering customer bases as it can be leveraged to predict churn and possibly apply suitable methods to circumvent potential user losses.
We train and test off the shelf machine learning techniques with 5-fold crossvalidation to predict user lifetime as a regression and classification problem; identifying the Random Forest to provide very strong results.
Discussing model complexity and quality trade-offs, we also dive deep into a time-dependent feature subset analysis, which does not work very well; Easing up the classification problem into a binary decision (lifetime longer than timespan $x$) enables a practical lifetime predictor with very good performance.
We identify implicit similarities across community models according to strong correlations in feature importance.
A single countrywide model generalizes the problem and works equally well for any tested community; the overall model internally works similar to others also indicated by its feature importances.
\end{abstract} 
\maketitle


\section{Introduction}

Every social networking platform depends on an active user-base.
This user-base is threatened by \emph{user churn}, which represents users leaving the platform.
Retaining existing users is a core marketing strategy~\cite{kotler2016framework} to mitigate potential losses focusing on positive user relationships via data and behavioral analysis.
Loyal users (possibly inadvertently) advertise a product freely.
More importantly, they tend being more profitable to a company.
Beyond our field, customer lifetime value (CLV) denotes expected revenue over time in marketing and may be used to identify high-value and users at risk.

The actual churn prediction's goal is not only limited to predicting a churn event, but also the likelihood or time until a user might churn.
Such individual churn probabilities allow for direct timed steering of single users (help, notifications, email) to improve individual retention.
In the broader picture, this also allows for steering communities or customer-bases with the envisioned optimal features in mind:
Maintaining a healthy user-base, which always is hoped to grow and converge into a well-mixed population.
The prediction of user churn is a well studied data mining task (cf. Related Work~\sref{sec:RW}).
However, these works predominantly focus on predicting churn in a single community typically represented by one platform.
The degree to which they generalize beyond a single user-base is thus an open question.
New types of location-based networks enable the establishment of many different disjoint communities within the same platform, e.g., enabling the study of information diffusion~\cite{Reelfs2019}.
This location-based property forming multiple independent user bases within the same plattform constrains thus enables the study of this currently open question on how churn models generalize beyond a single community.

In this paper, we analyze and predict user lifetime in Jodel, a mobile only \emph{location-based} social media messaging app.
Jodel establishes local communities relative to the users' location, thereby creating a multitude of individual local communities throughout a country.
As users cannot communicate with others outside their community, their user-bases tend to be disjoint.
This makes Jodel an interesting platform to study user lifetime in a large set of disjoint user groups / communities that are subject to the same application constraints while being comparable.
Further, current user churn studies focus on non-anonymous networks for which social ties often are a contributing factor (cf.~\sref{sec:RW}).
In turn, due to its anonymity Jodel enables us to study factors of user lifetime in the absence of such social ties or social credit.
Our contributions are as follows.
\begin{itemize}[noitemsep,topsep=5pt,leftmargin=9pt]
	\item We study user lifetime in a location-based, anonymous social messaging application.
	Our goal hereby is creating prediction models for the lifetime of a user within a specified observation period.
	We leverage resulting models to implicitly show in-/equalities of these communities \wrt{} churn.

	\item Among tested off the shelf machine learning algorithms, Random Forest provides the best results for predicting a users' expected lifetime, both in the case of a regression problem \emph{and} a classification problem.
	Our models use two types of features: user and community.
	We observe the models to perform well for all communities.

	\item Creating a single countrywide model generalizes the problem and works equally well for any tested community; this overall model internally works similar to others as indicated by its feature importances.
	We argue that model feature importances can provide feedback for empirical patterns pictured by the envisioned ideal community and may help to better understand reasons for users to stay or leave a platform.

	\item At last, we use Random Forest to answer a supposedly simpler---and easier to answer---binary classification problem of practical relevance to network operators: Given an observation time period, will the users' lifetime be longer?
	This approach achieves even better prediction quality than any other presented classifier.
\end{itemize}

\section{The Jodel App}

\begin{figure}[htb]
	\centering
	\includegraphics[width=0.99\linewidth]{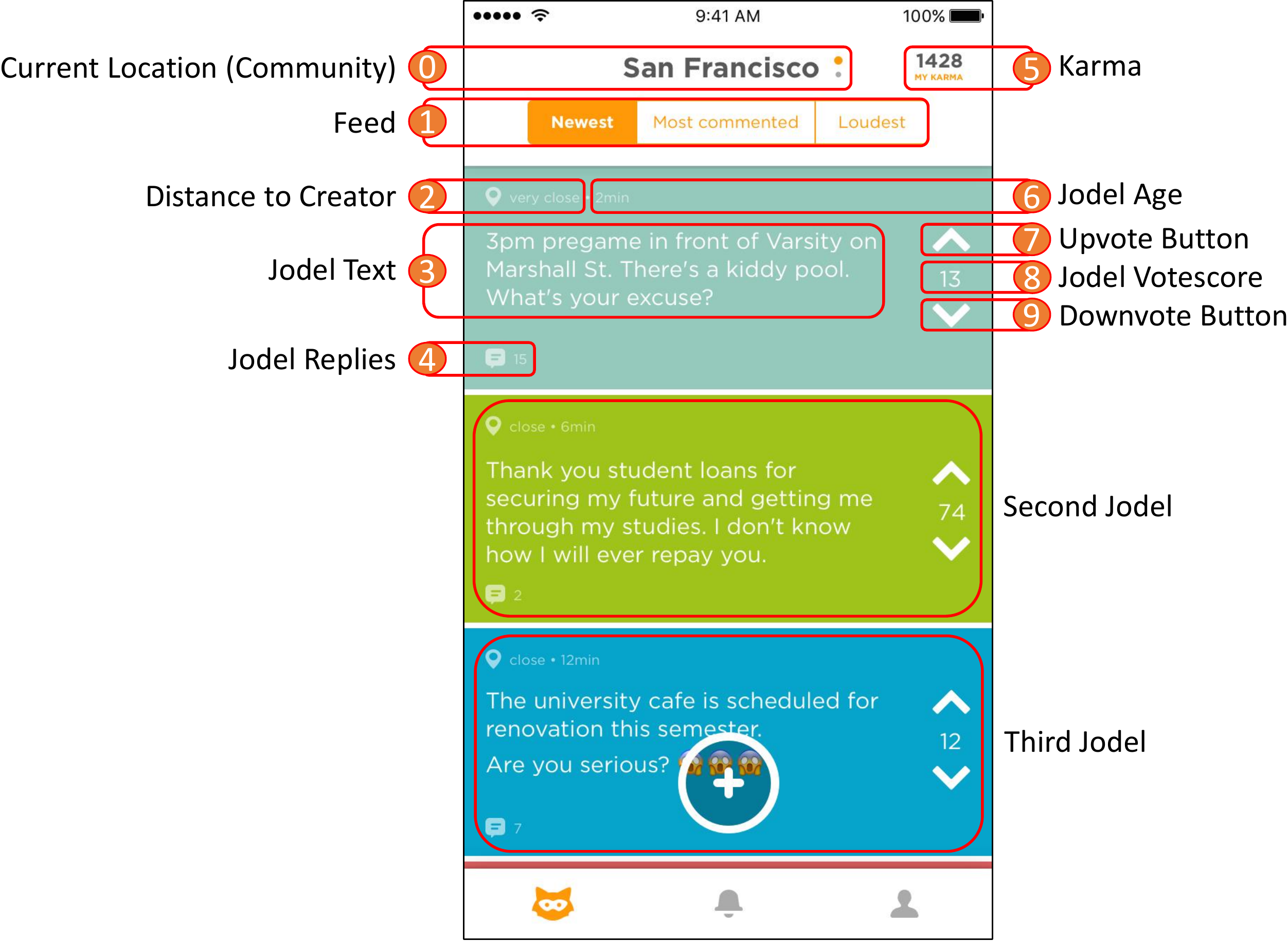}
	\caption{
		\textbf{Jodel iOS mobile application.}
	}
	\label{fig:jodelapp}
\end{figure}

Jodel\footnote{Jodel, German for yodeling, a form of singing or calling. ``Yudel'' \mbox{(\RL{يودل})} represents the adopted translation of Jodel to Arabic.}
is a mobile-only messaging application which we show in Fig.~\ref{fig:jodelapp}.
It is location-based and establishes local communities relative to the users' location \protect\encircle{0}.
Within these communities, users can {\em anonymously} post both images and textual content of up to 250 characters length \protect\encircle{3} (\ie{} microblogging) and reply to posts forming discussion threads \protect\encircle{4}.
Posted content is referred to as ``Jodels'', colored randomly \protect\encircle{3}.
Posts are only displayed to other users within close (up to $\approx$20km) geographic proximity \protect\encircle{2}.
Further, all communication is {\em anonymous} to other users since no user handles or other user-related information are displayed.
Only {\em within} a single discussion thread, users are enumerated to enable referencing to other users.
Up to 1.500 threads are displayed to the users in three different feeds \protect\encircle{1}: i) \emph{recent} showing the most recent threads, ii) \emph{most discussed} showing the most discussed threads, and iii) \emph{loudest} showing threads with the highest voting score (described next).

Jodel employs a community-driven filtering and moderation scheme to avoid adverse content.
For an anonymous messaging app, community moderation is a key success parameter to prevent harmful or abusive content.
In Jodel, content filtering relies on a distributed voting scheme in which every user can increase or decrease a post's vote score by up- (+1) \protect\encircle{7} or downvoting (-1) \protect\encircle{9}, \eg{} similar to StackOverflow.
Posts reaching a cumulative vote score \protect\encircle{8} below a negative threshold \mbox{(\eg{} -5)} are not displayed anymore.
Depending on the number of vote-contributions, this scheme filters out adverse content while also potentially preferring mainstream content.
Further, every post can be flagged as abusive, which subsequently is displayed to voluntary, system-selected, community moderators that majority-vote to remove or keep the particular post.
To increase overall user engagement in terms of creating content and voting, the network applies lightweight gamification by awarding \emph{Karma} points \protect\encircle{5}.

\section{User Lifetime}
\label{sec:churn}
To study and model user lifetime, we first describe our dataset and then define how we assess a user's lifetime.
The goal of our work is to automatically detect and classify the user lifetime in the anonymous and local communities within Jodel.

\subsection{Dataset \& Ethics}
\label{sec:Dataset_Description_and_Statistics}
The Jodel network operators provided us with {\em anonymized} data of their network.
The obtained data contains post and interaction \emph{metadata} for 469M posts/replies created within the KSA and spans multiple years from from 2014 up to August 2017, while within KSA the application has experienced reasonable usage only since late 2016.
For ethical reasons, it is limited to metadata only without textual content and user records stripped and anonymized.
The data enables us to analyze individual users' interactions by their anonymous ID.
Further, it contains no personal information and cannot be used to personally identify users.
We inform and synchronize with the Jodel operator on analyses we perform on their data.
The structure of our available dataset includes three object categories: interactions, content, and users.

\begin{itemize}[noitemsep,topsep=5pt,leftmargin=10pt]
\item
    \textbf{Users} (about 1\,M records) contains a user's accumulated karma value and whether the user is blocked.
\item
    An \textbf{interaction} (about 1\,B records) can be a \emph{registered}, \emph{post}, \emph{reply}, \emph{upvote}, \emph{downvote}, or \emph{flag}.
    Each interaction has a timestamp and a geohash.
    It is further linked to a user ID and a content ID.
\item
    \textbf{Content} (about 469\,M records) may either be a new post, \ie{} starting a new thread, or a reply.
        This content records include a boolean flag whether it is text or an image (note: videos were added after our observation period).
\end{itemize}

\afblock{Dataset limitations.}
Our dataset only includes the users' {\em active} interactions with the system, \ie{} registering, creating posts, replying, or voting.
Thus, we cannot infer when or how much a user only {\em passively} participates---lurkers---who only consume content.
Further, the vote interactions are always mapped to the date and geoposition of the respective content creation.
This prevents us from making detailed analyses depending on the voting time or place.
However, due to the vivid usage of the application within larger communities (multiple posts/replies per minute), we generally consider votes to be executed on the same day as their respective content.
Especially since posts are only accessible via the three different feeds, where they will only stay for a very limited time, casting votes to content long after creation is usually not possible.

\subsection{Goal: User Lifetime Prediction}
\label{sec:churn_definition}

\afblock{Definition.}
We define the \emph{lifetime} of a user as the time between the first (automatic account creation) and last system interaction (\ie{} posting or voting) in minutes.
Note that we can only define the lifetime by system interactions of a user since our dataset does not include passive activities (\ie{} only reading).
In prior work, churn, \ie{} users leaving the system, is often defined as the end of a user's lifetime.
The lifetime enables us to partition users into timespan-dependent activity classes that we later predict, \eg{} users that only used Jodel for a short amount of time, or longer.

\begin{table}[tbh]
    \begin{tabular}{clrr}
        \toprule
        \textbf{class} & \textbf{lifetime} & \textbf{\# users} & \textbf{fraction} \\
        \midrule
        1 & $0 \ldots 1$ days &     $135$k & $13.3\%$\\
        2 & $1 \ldots 7$ days &     $123$k & $12.1\%$\\
        3 & $7 \ldots 14$ days &    $75$k & $7.4\%$\\
        4 & $0.5 \ldots 1$ month &  $124$k & $12.3\%$\\
        5 & $1 \ldots 3$ months &   $268$k & $26.4\%$\\
        6 & $> 3$ months &          $288$k & $28.4\%$ \\ 
        \hline
        $\Sigma$  &          & $1\,012$k      & $100.0\%$ \\
        \bottomrule
    \end{tabular}
    \caption{The definition of six churn classes.
        We subdivide the user population by their active time.}
    \label{tab:classes}
\end{table}

\afblock{User lifetime distribution.}
We compute the lifetime for every user in our dataset and group users into six classes shown in Table~\ref{tab:classes}.
There is a broad range of different user lifetimes, ranging from users that only use Jodel for less than a single day and users that stick with the platform for more than 3 months.
A user's lifetime can end in two cases:
First, the end of the observation period.
Second, users that stop using the application (\ie{} churn).
As in a practical setting the observation period is always finite, prior work \emph{approximates} the churn potential by using a threshold (\eg{} no activity within the last $n$ days, where $n$ often is derived empirically); the finiteness naturally introduces a skew towards shorter time periods.
If we apply such a threshold of one week $n=7$ (\ie{} users are regarded as churners if there is no activity within the one week threshold margin towards the end of observation), 61\% of the users will be defined as churners.
We remark that churn prediction is an inherently hard problem since users could become active again after the threshold.
Instead of churn, we predict user lifetime, \ie{} the chance of a user to use the app for at least $k$ days or a discretized timespan.

\afblock{Paper goal.}
The goal of the paper is to predict the lifetime of a user within a specified observation period.
Social network operators can use the resulting models as an online algorithm to predict the likelihood of a user to stick with the platform; furthermore, they allow for studying user behavior.
\section{Features}
\label{sec:features}

To predict the lifetime of a user by using a data-driven ML model, we derive features from \emph{i)} the user itself and \emph{ii)} her community.

\afblock{Engineering Subsets.}
We introduce two different feature classes to represent the user and her environment:
\emph{i)} \textbf{User} related \textbf{features}: \eg{} user registration event information, down-/upvoting and post/reply behavior of a specific user.
\emph{ii)} \textbf{Community} related \textbf{features}: \eg{} posts/re\-plies, up-/downvotes, average post response time of the users home community.

\afblock{Users home community.}
Since Jodel establishes communities relative to the users' locations, users can participate within different communities when sufficiently changing their geographic position.
For a stable model, we derive the community features from the users' home community, which we define as the city location with a user's most interactions.
For $87\%$ of the users, this home community represents the city in which they initially registered.
We use this attribute to determine a user's city throughout this work.

\afblock{Capturing time.}
While our features up to now do not catch any time-dependent information reflecting a user's lifecycle, we add timing insights by duplicating the features with time-period bounds (1 day, 2 days, 3 days, 1 week, 2 weeks, 1 month, 3 months, $>$3 months).
We end up with 66 (29) time-(in)dependent features.
Noteworthy, none of typical scaling (\eg{} std, min-max) or balancing (\eg{} SMOTE, random over/undersampling) techniques improved any of our results significantly.
Further, we used mean imputing as it provided best overall results.
\section{Random Forest Modelling}
\label{sec:ML}
Within this section, we discuss our machine learning approach for a user lifetime predictor.
By applying grid searches, after showing the Random Forest providing best results in comparison to other off the shelf ML methods, we dive deeper into the results of individual community models and their best parameters obtained.
Then, we show how the prediction quality depends on model complexity and how different time-dependent feature subsets determine performance.
Moreover, we look into model generalization and the impact of the amount of input data.
Eventually, we derive implicit model (and as such community) similarities. %

For any model, we apply grid-searches levraging a random 5-fold crossvalidation approach for both, formulating a problem for \emph{a)} regression of lifetime in minutes (REG), and \emph{b)} classification into the previsouly defined classes (CLF).
Our main evaluation metrics are: \emph{REG:} The $R^2$ score measures how equal real and predicted populations are, and \emph{CLF:} the F1 score describes the harmonic mean of precision and recall; both providing an overall picture. %

Since the property of establishing communities relative to a user's position, our data set contains a large set of city-level communities throughout the country.
To focus our discussion, in comparison to an all data \emph{Country} model, we chose a subset of five cities varying in their size by the amount of users to create distinct prediction models: Riyadh, Jeddah, Mecca, Al Bahah, and Al Jafr (large to small).
Our implementation uses Python skicit-learn off the shelf functionality.

\subsection{ML Algorithm Selection}
First, we ran a grid search for all data over a set of hyperparameters for standard ML methods to obtain ballpark numbers.
These grid searches use a \emph{mean} imputing strategy while not incorporating any scaling or balancing.
Our used algorithms are Random Forest (RF), Decision Tree (DT), Multi Perceptron (MLP), AdaBoost (AdaB), K-Nearest Neighbors (KNN), and Stochastic Gradient Descent (SGD). %

Table~\ref{table:algorithm_results} provides an overview of a baseline in comparison to both problem formulations: regression and classification.
The baseline is obtained by using the scikit-learn dummy regressor predicting the mean user lifetime in minutes, and the most frequent class for classification.
While it yields unusable results for the regression task, an imbalanced dataset naturally provides better figures for classification.
Due to multiple crossvalidation runs, we also provide standard deviation figures.

\begin{table}[tbh]
    \begin{tabular}{lrr}
        \toprule
        Algorithm & REG: $R^2$ $\pm$ stddev & CLF: F1 $\pm$ stddev\\ \hline
        \midrule
        baseline & $-0.0000 \pm 0.0000$ & $0.1672 \pm 0.0007$ \\ \hline    
        RF & $0.9822 \pm 0.0004$ & $0.9668 \pm 0.0005$ \\ 
        DT & $0.9580 \pm 0.0009$ & $0.8049 \pm 0.0095$ \\
        MLP & $0.9668 \pm 0.0036$ & $0.6768 \pm 0.0304$ \\ 
        AdaB & $0.7654 \pm 0.0012$ & $0.6720 \pm 0.0093$ \\ 
        KNN & $0.6764 \pm 0.0013$ & $0.5077 \pm 0.0006$ \\ 
        SGD & $0.3422 \pm 0.0551$ & $0.1686 \pm 0.0191$ \\ 
        \bottomrule
    \end{tabular}
    \caption{Off the shelf ML algorithm results using all data applying mean imputing; no scaling, no balancing.
        While the RF performs best, DT and MLP achieve similar regression performance falling short in classification. %
        }
    \label{table:algorithm_results}
\end{table}

We observe that both, the regression and classification baseline are easily outperformed by any algorithm except for CLF with SGD and KNN.
The best performing algorithm always is a rather complex Random Forest. %
However, for regression, DT and MLP also perform quite good. %
The results almost do not fluctuate across multiple crossvalidation instances at all.

\takeaway{The best performing ML algorithm is the Random Forest with very strong regression $R^2\approx 0.97$ and classification $\text{F1} \approx 0.99$ scores.
Thus, we will from now on focus on the RF algorithm.
Nonetheless, most others also outperform the baseline significantly.}

\subsection{Independent Communities}
\label{sec:community_results}
As we have now determined the best-working algorithm for all data to be Random Forest and its parameters for our regression and classification predictor, we take a closer look into performance of specific independent community models.
That is, does the prediction performance differ by community?
This evaluation is enabled by the location-based nature of Jodel which allows us to compare independent user bases subject to the same platform constraints.

In Table~\ref{table:city_comparison}, we show the best results of each Random Forest grid search instance for our selected cities and an all data country model.
The communities are sorted by the amount of users within their community in descending order.
We selected these particular examples due to their different amounts of users to cover a wide range from large to small.
The $R^2$ score describes the crossvalidation result for the regression problem, whereas F1 describes the classification results; additionally, we provide the standard deviation across folds.

\begin{table}[tbh]
    \begin{tabular}{lrrr}
        \toprule
        Community & REG: $R^2\pm$stddev & CLF: F1$\pm$stddev & \#users\\
        \midrule\hline
        County & $0.9822 \pm 0.0004$ & $0.9668 \pm 0.0005$ & $1\,012\text{k}$\\ \hline
        Riyadh & $0.9728 \pm 0.0013$ & $0.9531 \pm 0.0006$ & $284 \text{k}$ \\
        Jeddah & $0.9667 \pm 0.0016$ & $0.9372 \pm 0.0008$ & $101 \text{k}$ \\
        Mecca & $0.9551 \pm 0.0035$ & $0.9185 \pm 0.0039$ & $45 \text{k}$ \\
        Al Bahah & $0.9457 \pm 0.0032$ & $0.8752 \pm 0.0093$ & $11 \text{k}$ \\
        Al Jafr & $0.8219 \pm 0.1115$ & $0.5807 \pm 0.0594$ & $174$ \\
        \bottomrule
    \end{tabular}
    \caption{RF classification and regression results for the Country model and selected individual communities.
     Results are consistent and stronger for larger communties, except for Al Jafr due to small amount of data.
    }
    \label{table:city_comparison}
\end{table}

Generally, all predictions perform quite well with $R^2$ scores above $0.94$ for all cities except Al Jafr; The F1 score within the classification instances is above $0.87$ for most cases.
There are negligible fluctuations across folds as seen by the low standard deviation for all cities again except Al Jafr due to the very few data points (only 174 total users); yet, the regression still works surprisingly well, whereas the classification falls short in achieved quality.
We observe that all predictors work better on larger communities, hence more data. %

Further, besides our best performing Random Forests mostly use larger amounts of estimators, we notice that they are complex being rather deep and leveraging all features ($n$).
We will discuss tree complexity versus performance in Subsection~\ref{sec:model_sweetspot}.

\takeaway{The overall predictor performance for the regression and classification task is very good for all analyzed independent communities except Al Jafr.
Resulting models are quite complex in terms of tree depths, used features and estimators.
Best-performing classifications tend to require less complex model instances than the regression.}

\subsection{Predictor Sweet Spot}
\label{sec:model_sweetspot}
As we have presented best performing results from the grid search for the Random Forest across our selected communities, we now want to shed light on the relationship between model complexity and prediction quality for two reasons: 
\emph{1)} An overly complex model might tend to overfit our data, 
\emph{2)} Less complex models are usually preferred due to less computation times for both, fitting and application.

We therefore investigated the relationship between used features, estimators, tree depth and resulting model quality for both, CLF and REG.
For both problem formulations, model complexities are qualitatively very similar to their achieved performance---expectedly, more complex models perform better.

\begin{figure}[htb]
    \centering
    \includegraphics[width=\linewidth]{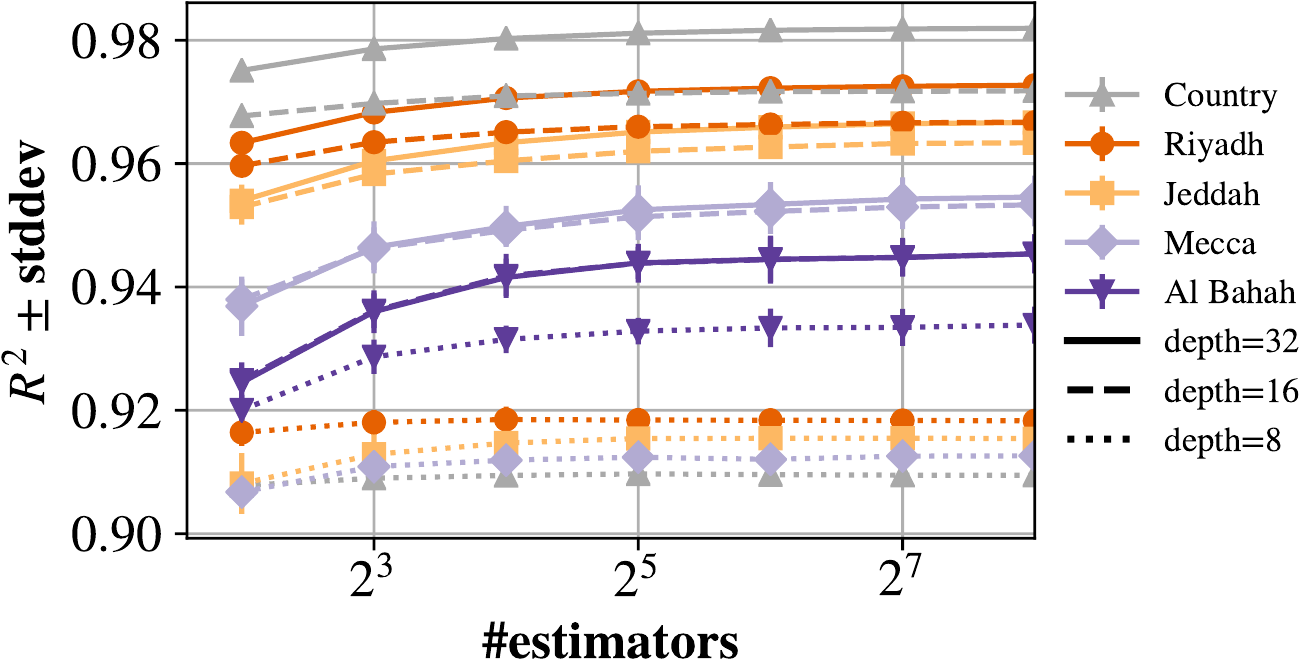}
    \caption{RF regression performance vs. model complexity.
        More complex models provide stronger performance with diminishing returns in depth and especially the amount of used estimators.
        }
    \label{fig:rf_parameters}
\end{figure}

Exemplary, we show the relation between used estimators, tree depth and quality for the regression model using all features in Figure~\ref{fig:rf_parameters}, which ultimately allows us to define a sweet spot.
The logarithmic x-axis denotes the amount of used estimators (tree instances of the ensemble), whereas the y-axis shows the resulting $R^2$ score and standard deviation as error bars.
There are three series for each city for a tree depth of 8, 16, and 32, respectively; we removed Al Jafr to increase readability.

We observe that increasing the tree depth substantially increases model quality.
However, the improvement from a depth of 8 in comparison to 16 is by far larger than the change from 16 to 32.
For the smallest community shown, Al Bahah, increasing the tree depth above 16 does not improve performance.
While 32 estimators already yield very good results, the quality increase is asymptotically bounded, \ie{} there are diminishing returns.

\takeaway{Although our grid search shows best results with rather complex Random Forest models (cf.~\sref{sec:community_results}), our in-depth analysis of hyperparameters vs. quality reveals that only few estimators with mediocre tree depths already yield very good results with diminishing returns of increasing model complexity.}

\subsection{Feature Subset Analysis}
\label{sec:feature_subset_analysis}
While our prediction for regression and classification works well, we next want to determine the impact of different feature subsets.
That is, which feature subset provides best results, or even better results than using all features?
We conduct grid searches for the Random Forest across community and user features for all data (Country) and each of our selected independent communities.

Furthermore, from a practical implementation standpoint, one might think about leveraging a sliding window approach over a single day, a week or longer as model input.
Such an approach makes a model more time-invariant.
However, limited knowledge may seriously impact the model performance---as it is expected to degrade.

\begin{figure}[t]
    \includegraphics[width=\linewidth]{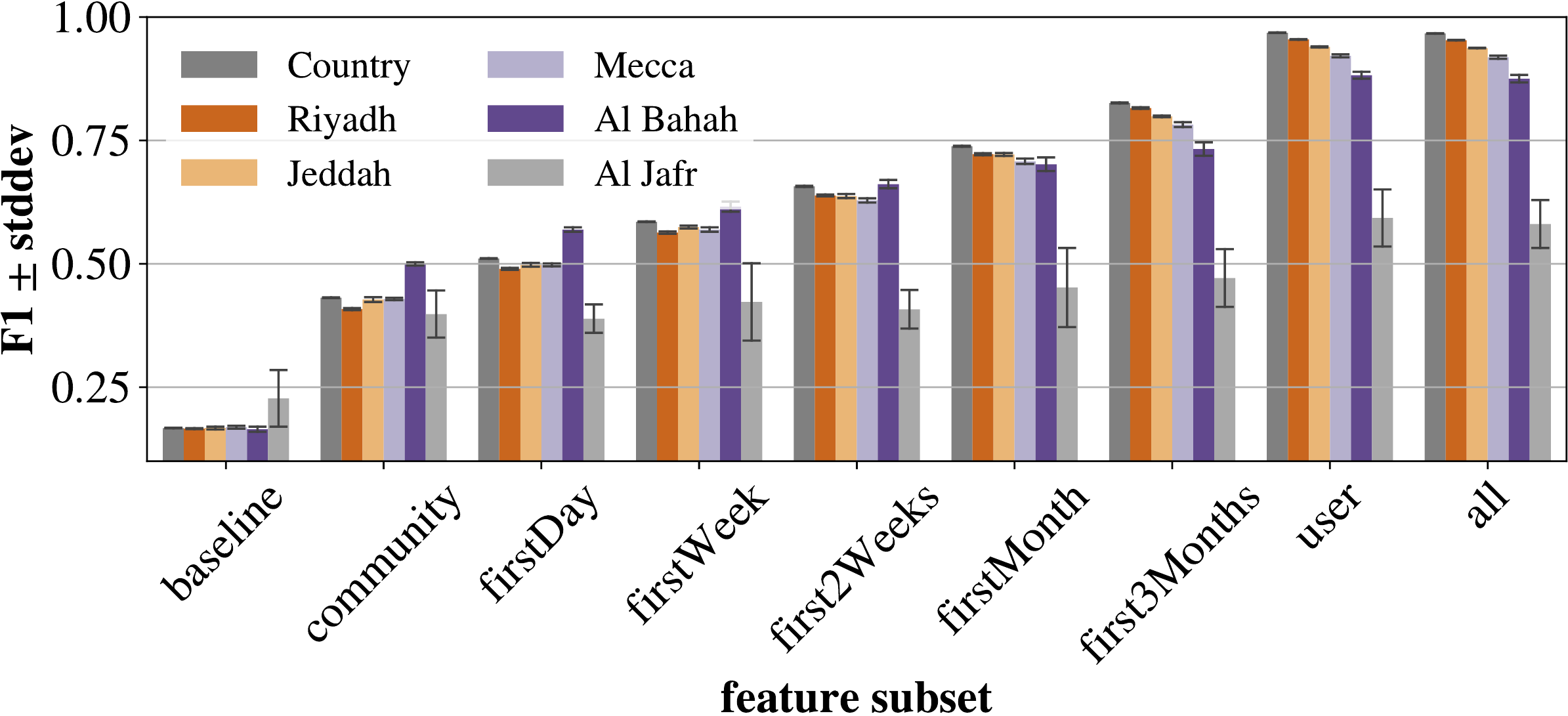}
    \caption{RF Classification feature subset performance results.
        With increasing observation time per user, the results improve significantly.
        Overall, the country model works best, while most others achieve similar performance.
        }
    \label{fig:feature_subsets}
\end{figure}

Thus, we show the impact of our feature subsets within Figure~\ref{fig:feature_subsets} on the classification example.
Note that the subset impact is quite similar for the regression problem (not shown).
The x-axis describes used feature subsets: community and time-dependent features being cumulative: \Eg{} the \emph{firstWeek} subset also includes features for the shorter time periods of \emph{firstDay} and \emph{first3Days}.
The y-axis denotes the model quality via the F1 metric and the standard deviation across folds.

First of all, we observe that the community features alone provide worst results, but are quite similar to only looking into user data of her individual first days at scores ranging from $\approx 0.4 \ldots 0.6$.
By increasing the observation time-window up to 3 months, the model quality increases drastically to F1 scores $>0.8$ for most cities.
Only relying on user features provides similar results to using all features; the community features by itself only have a negligible impact on prediction quality.
To be clear, this does not imply that the community only has negligible impact on user lifetime or user experience. %

Noteworthy, predictions for larger communities tend to be better than for \eg{} the Al Jafr community, always being off presumably due to its very few users (only 174).
We cannot explain why the predictions for Al Bahah perform better for short time frames.

\takeaway{Although our prediction models have proven very good performance, taking a practical stance by only using timely-windowed features depending on the users' active time reveals that classification and regression (not shown) quality deteriorates for real-world use-cases.}

\subsection{Generalization} %
We next study how well our models predict the lifetime from other communities to investigate whether there is a model---possibly of a community, or the County-model---providing well-suited prediction quality that may be used as a single all-round country-wide model.

\begin{figure}[ht]
    \includegraphics[width=.7\linewidth]{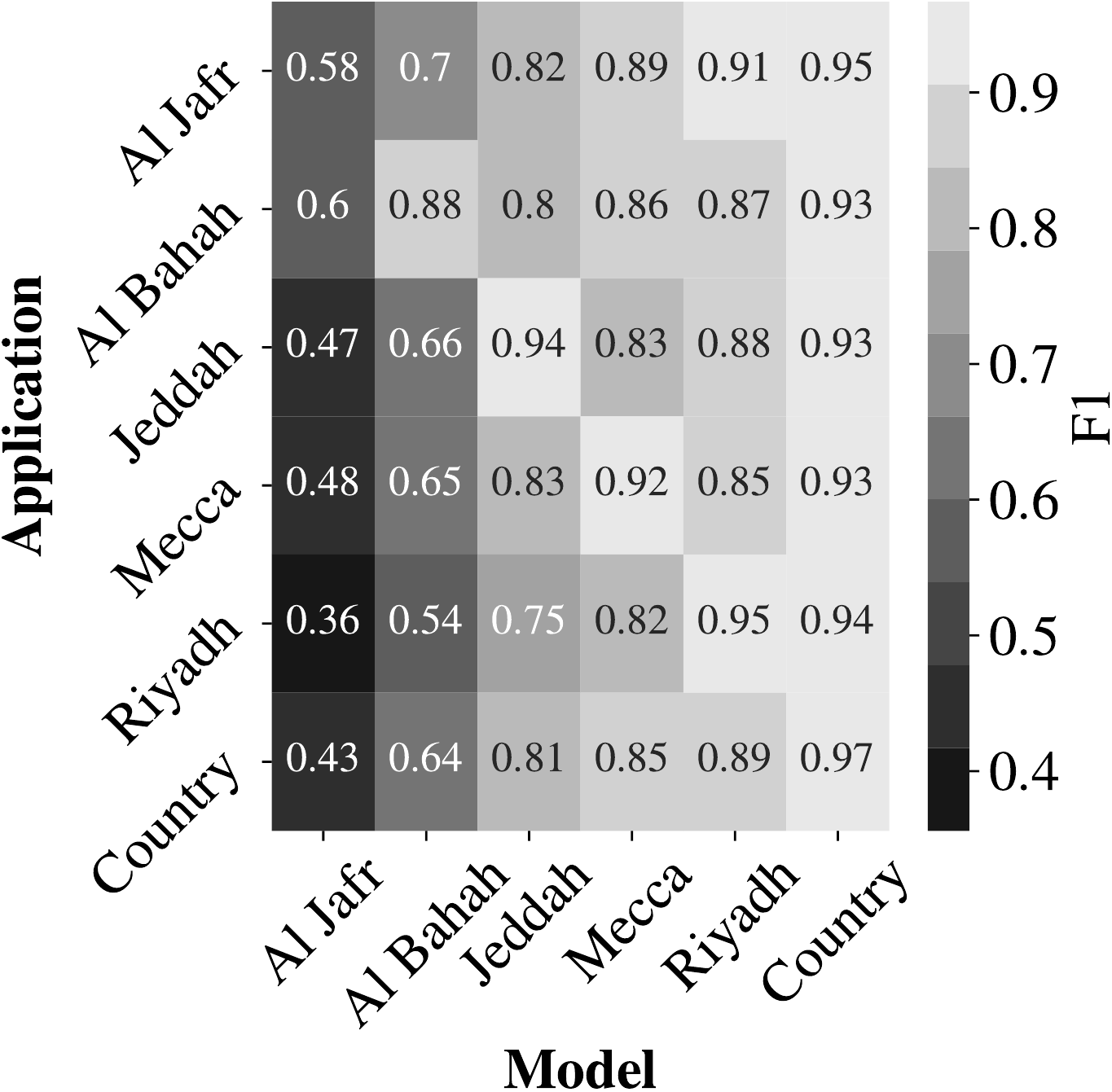}
    \caption{RF classification model cross application results.
        We used each created model and performed a prediction for every other dataset.
        The diagonal same-same instances depict earlier prediction results as a comparison.
        The country model works best across the board.}
    \label{fig:cross_application}
\end{figure}

Within Figure~\ref{fig:cross_application}, using the best-performing model from our previous grid-search (cf.~\sref{sec:community_results}) each, we provide cross-application (community-model $\times$ predictions-for-community) classification scorings of our different community and the Country model(s).
The x-axis describes the used model instance, whereas the y-axis denotes the predictor input-dataset.
We provide the macro F1 scores for each combination, colored on the z-axis.
Note that we added the same-same community model/application F1 scores from previous results as a baseline (diagonal upper left to bottom right).

Focusing on the \emph{County}-model first, \emph{i)} we observe that it provides strong generalized performance with F1 scores $\geq 0.93$ throughout any community (rightmost column).
Taking a closer look into the community cross applications, we find:
\emph{ii)} Most individual community models perform very well on their own input dataset; other models significantly improve the Al Jafr community prediction scores.
\emph{iii)} The overall best-working community model is Riyadh, falling short in prediction quality for Mecca.
The community models from Mecca (Jeddah) still delivers acceptable prediction quality across the board with F1 scores $\geq 0.82$ ($\geq 0.75$).
However, the models for the smaller communities do not perform well in a generalized setting.
\emph{iii)} By leveraging the model scores as a proxy for community similarly \wrt{} user lifetime, we identify that due to shown options for generalization, user lifetime is similar to some extent across the analyzed communities, yet through the models only being implicitly defined. %
\emph{iv)} For the regression use-case, we find the Country model strictly outperforming all others with an $R^2$ score of $0.98$ in all communities; besides, only the Mecca community model provides consistent strong results applied to other communities at $R^2$ scores $\geq 0.83$ (not shown).

This generalization reveals that the independent communities can be captured well in a single model and behave similar to some extent \wrt{} user lifetime.
Note however, the \emph{Country}-model may be skewed in favor of larger communities due to their heavier impact within the imbalanced dataset.

\takeaway{The overall Country model performs very well throughout any tested community for both, regression and classification, with improvements for smaller communities and (for classification) slight deteriorations for the clique of larger cities (Riyadh, Jeddah, Mecca).
Yet this model works well and might be used for the whole dataset as a unique predictor, retraining is computationally heavier than selected individual models due to its size.
}

\subsection{Country Model in Detail}
Our evaluation and cross applicaton showed that the \emph{Country}-model provides all-round performance for regression and classification, while also improving predictions for communities with comparably fewer users.
But why is that?
Does simply the amount of available data improve the model, or does the country model represent a better cut through the population?

To answer this question, we randomly downscaled all data to the reference values of our other selected communities and ran grid-searches for these new sampled Country models.
We present our results in Table~\ref{table:country_detail} for the classification problem. %
The alike community column depicts the reference sample size, whereas the F1 country column denotes the model's classification quality.
Further, we add the individual city models as an expected upper-bound baseline comparison (column F1 city).

\begin{table}[tb]
    \begin{tabular}{llrrr}
        \toprule
        \# &     alike &  F1 country $\pm$ std & F1 comm. $\pm$ std&   $\Delta$ \\
        \hline\hline
        0 &  Country &              $0.9668 \pm 0.0005$ & -                   & -\\ \hline
        1 &   Riyadh &              $0.9496 \pm 0.0016$ & $0.9531 \pm 0.0006$ & $-0.0035$ \\
        2 &   Jeddah &              $0.9312 \pm 0.0026$ & $0.9372 \pm 0.0008$ & $-0.0060$ \\
        3 &    Mecca &              $0.9136 \pm 0.0024$ & $0.9185 \pm 0.0039$ & $-0.0049$ \\
        4 & Al Bahah &              $0.8773 \pm 0.0063$ & $0.8752 \pm 0.0093$ &  $0.0021$ \\
        5 &  Al Jafr &              $0.6794 \pm 0.0691$ & $0.5807 \pm 0.0594$ &  $0.0988$ \\
        \bottomrule
    \end{tabular}
    \caption{RF classification Country model with limited input data corresponding to other city sizes.
        The country model achieves similar performance to individual models when restricting the input data size.
        \Ie{} by reducing the amount of data, performance deteriorates.}
    \label{table:country_detail}
\end{table}

Our evaluation reveals that the model performance remains very strong, but worsens with less data.
However, given a statistical significance due to the sheer amount of input data, the observed delta to the individual model arguable remains within margin of error; except for Al Jafr most probably due to its very sparse data.

\takeaway{We observe that the country model performance is similar to the individual models provided it uses the same amount of input data.
Still, the country model deals better throughout all communities than any other individual model.
}

\subsection{Feature Relevance per Community}
\label{sec:beyond_ml}
We observe the Random Forest to provide good prediction results.
Futher, we have seen that the communities behave similar \wrt{} user churn to some extent.
However, we are missing information which metrics were predominantly used by the models.
Do they rely on the very same features or have they learnt differently?

\begin{figure}[htb]
    \includegraphics[width=.7\linewidth]{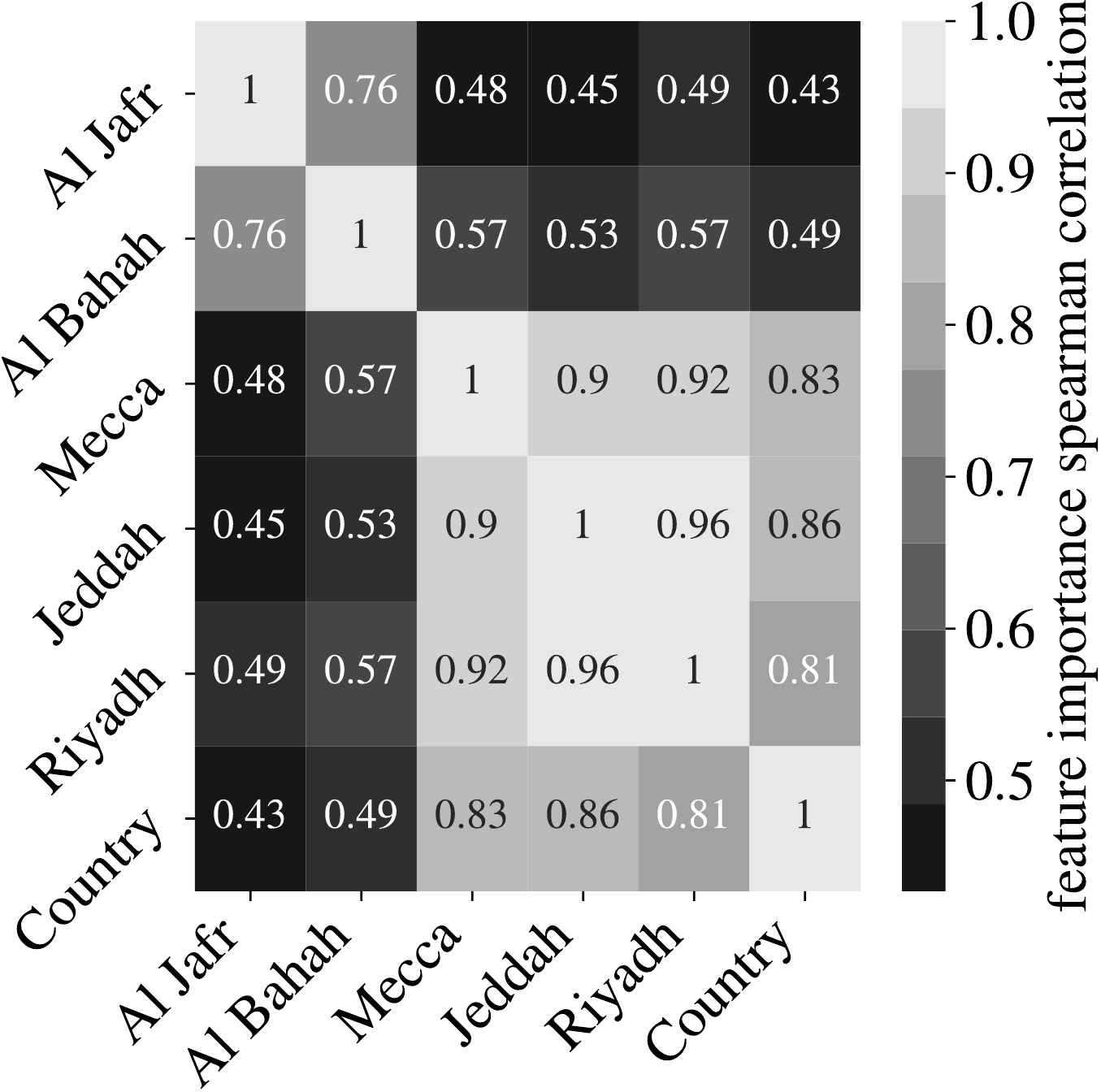}
    \caption{RF classification features importance spearman correlation.
        We find that most models strongly correlate in feature importances indicating similar model internals.}
    \label{fig:correlation_rf_importance}
\end{figure}

There are different well-known feature importance predictors, such as ReliefF, RFECV or the input variance.
However, we focus on RF Feature Importance (Gini importance/Mean Decrease in Accuracy), which depict the percentage ranking within its decision making.
By cross correlating the feature importances with the ranked Spearman's, we next want to find out how well the importance rankings across the different models line up.
We show the results for the classification problem in Figure~\ref{fig:correlation_rf_importance}, which depicts a community on both axes, whereas the ranked correlation coefficient (Spearman) between the cities' importance vector is represented textually on the also colored z-axis.
Note that the overall picture remains the same for Regression (not shown).

We observe that the clique of Mecca-Jeddah-Riyadh line up quite well, whereas the smaller cities fall off.
Interestingly, the Country model feature importances also correlate to beforementioned clique indicating that these models work similarly.
Digging deeper into the Country model by correlating importances of the sampled versions (not shown), we find that all of these models, despite using less data, are strongly correlating in feauture importance and thus, share similar internals; except alike-Al Jafr falling short.

\takeaway{
    The rather strong correlation among the models for larger cities strengthens our hypothesis of similarly behaving communities due to their model similarities.
    This also holds true for the overall Country model and its sampled versions.
}

\subsection{Future Work: Empirical Lifetime Study}
\label{sec:future_work}
Having built well-performing predictors for user lifetime and having seen that model feature importances often correlate, these importances figure an important signal of usage within the respective model, which deserve to be fed back into a thorough empirical analysis to better understand model internals.

We argue that used feature values of users subdivided into their lifetime classes represent a projection of its population \wrt{} user lifetime.
\Ie{} by partitioning a feature's population by the optimization target (active time), we discretize the community state available to each user.
Implicitly driven through lifetime predicting RFs, a community state for important single features \wrt{} lifelime might then be given by the distribution in a certain time-slice---which in turn depict qualitative differences between these partitions.

However, while punctual information is not sufficient to identify changes, we are more interested in qualitative changes over time within most important features.
To provide empirical insights, we first removed upper and lower 1\% outliers and then applied MinMax scaling (onto 0 to 1).
Then we calculated the set of quantiles $10\%,20\%,\ldots,90\%$ and plotted the resulting kernel for each feature over the time subsets (1 day, 3 days, 1 week, 2 week, 1 month, and above) in Table~\ref{table:empirics}, while the solid line denotes the median.
That is, the (second) outmost gray area denotes the population between the quantiles of $10\%$ ($20\%$) and $90\%$ ($80\%$).
To better capture qualitative changes, we apply a logarithmic y scale.

\begin{table}[htb]
    \begin{tabular}{rccccccc}
        \toprule
        feature &  \,\,\, 1d &  3d &  1w &  2w &  1m &  3m &  < \\
        \midrule
          postcreated    & \multicolumn{7}{c}{\includegraphics[valign=m,width=5cm]{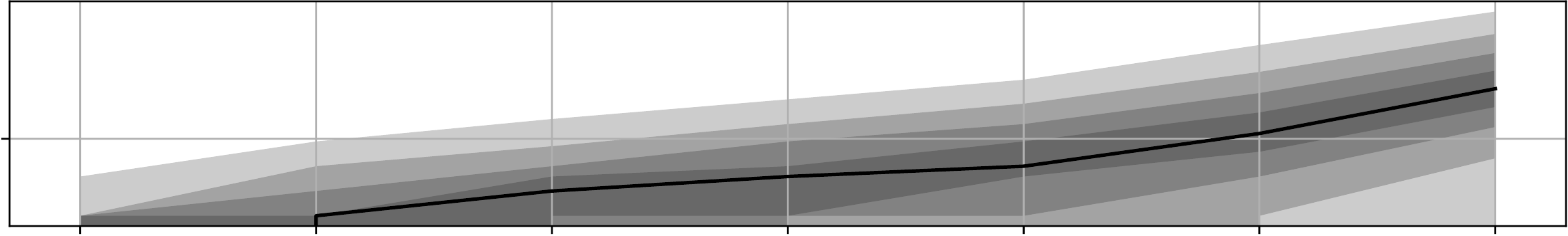}} \\
          min\_btwn\_posts    & \multicolumn{7}{c}{\includegraphics[valign=m,width=5cm]{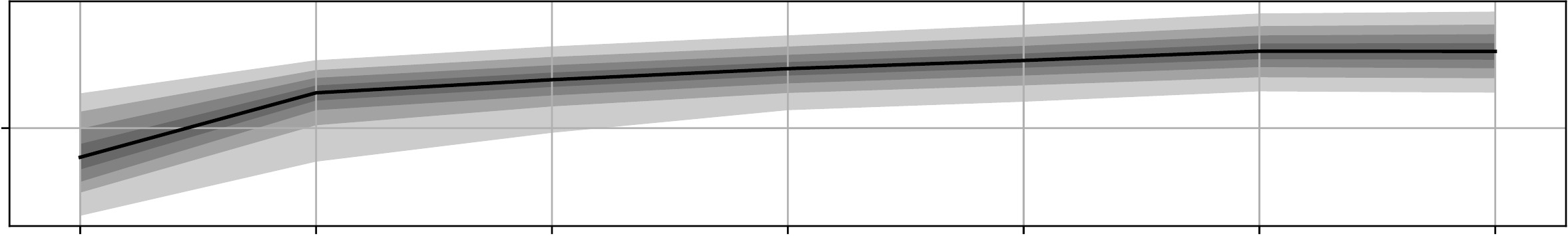}} \\
          replies\_day    & \multicolumn{7}{c}{\includegraphics[valign=m,width=5cm]{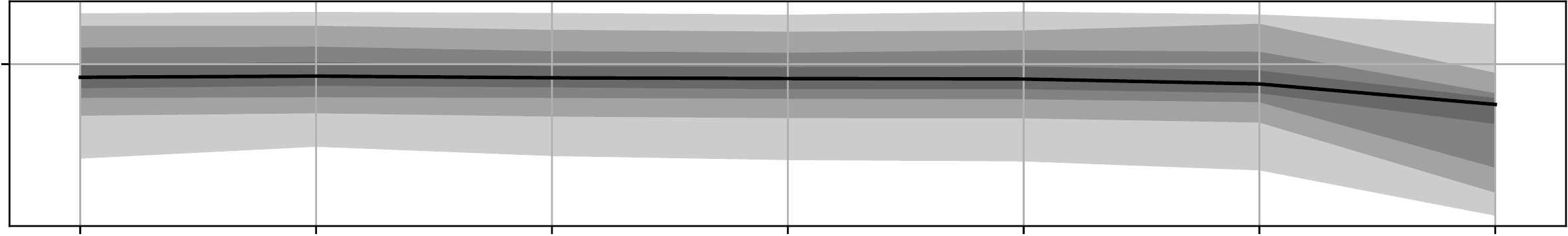}} \\
          RegPostGap\_h    & \multicolumn{7}{c}{\includegraphics[valign=m,width=5cm]{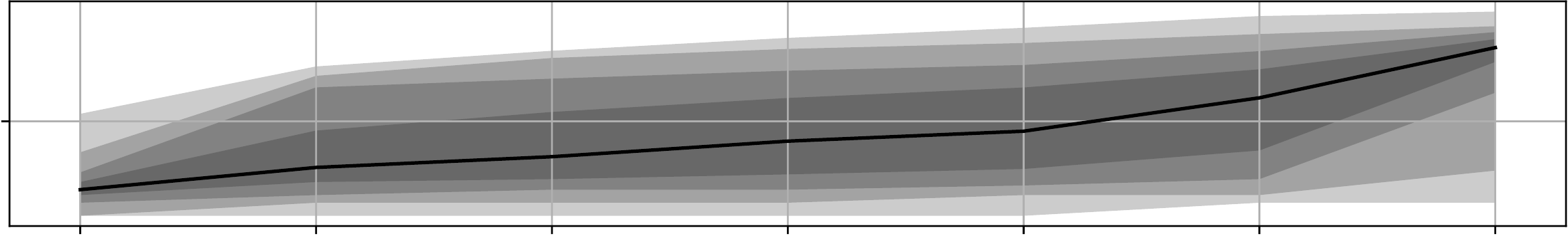}} \\
          min\_btwn\_intrctns    & \multicolumn{7}{c}{\includegraphics[valign=m,width=5cm]{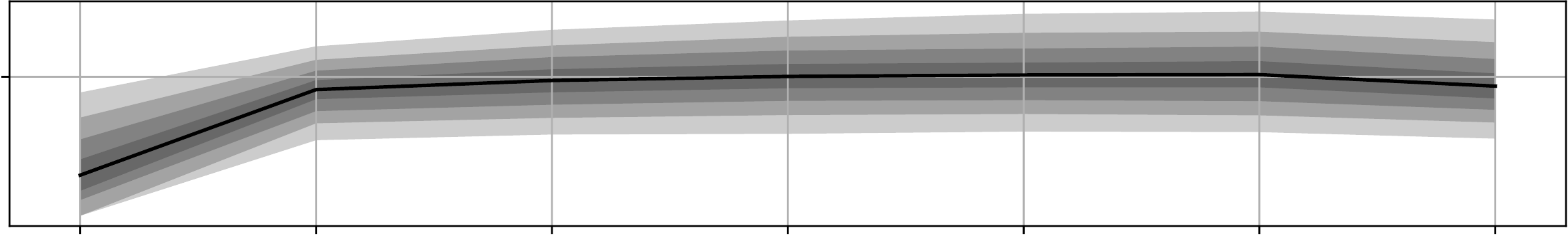}} \\
          replycreated    & \multicolumn{7}{c}{\includegraphics[valign=m,width=5cm]{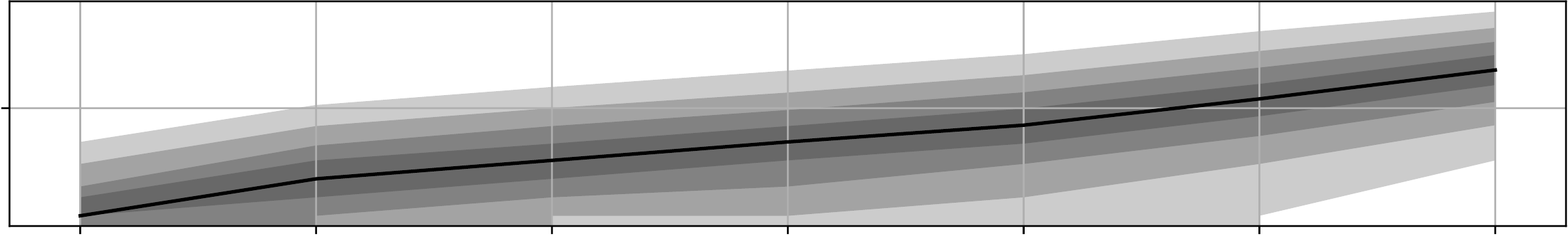}} \\
          picture\_posts\_day    & \multicolumn{7}{c}{\includegraphics[valign=m,width=5cm]{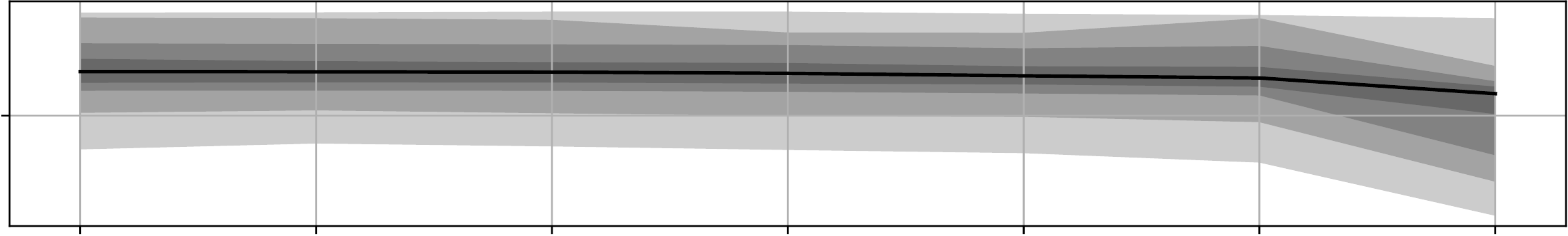}} \\
          replycreated\_day    & \multicolumn{7}{c}{\includegraphics[valign=m,width=5cm]{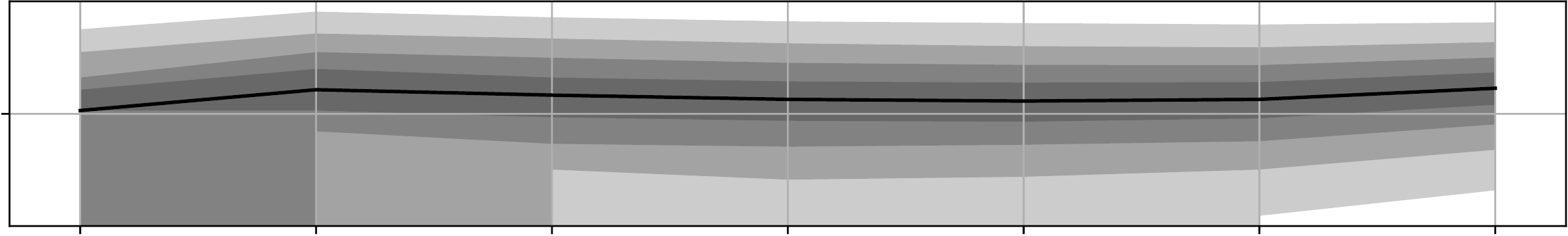}} \\
          downvotes    & \multicolumn{7}{c}{\includegraphics[valign=m,width=5cm]{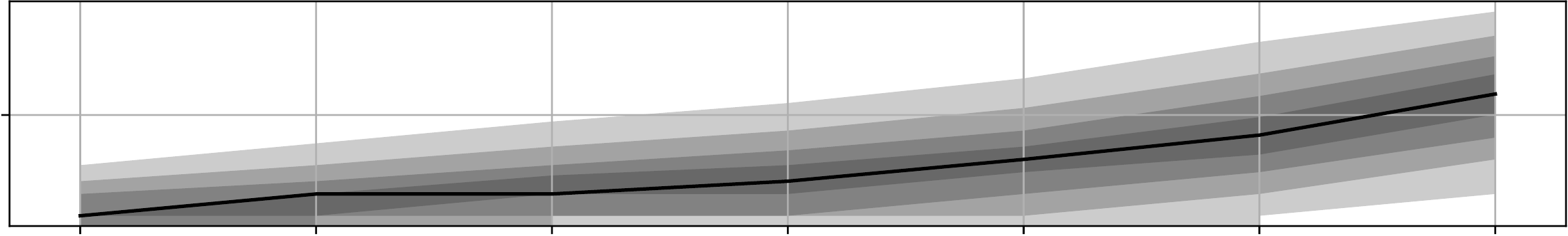}} \\
          \bottomrule
    \end{tabular}
    \caption{Qualitative population of most important features (country model) subdivided into user lifetime sets.
        The areas denote quantiles from 10-90\%, 20-80\% etc. whereas the solid line depicts the median.
        On a logarithmic y-axis, we observe drastic variations within the metrics over a user's lifetime.}
    \label{table:empirics}
\end{table}

Given new challenges like for some metrics, we can observe clear trends, while others draw an unclear picture, or any counting feature depending on time, we leave next steps for future work.
Nonetheless, feature candidates with clear trends might be a valuable object for future research, \ie{} learning from model behavior; and ultimately \eg{} testing hypotheses with \eg{} synthetic tests~\cite{Amjad2019}. %

\section{Binary Lifetime Prediction}
Although our prediction works quite well overall with classification F1 scores up to $0.95$ for Riyadh, having only timely limited user information deteriorates prediction quality significantly (cf.~\sref{sec:feature_subset_analysis}). %
In practive, \eg{} a network operator usually only asks whether a user is likely to churn in near future.
This allows us to reformulate our problem into a supposedly simpler--and easier to answer--binary classification problem:
Given an observation time period, will a user's lifetime be longer?

Thus, we ran grid searches for binary predictors for every lifetime class in Table~\ref{tab:classes} and all selected communities likewise to the feature subset analysis, which allows us to generate predictions according to our chosen time-periods of the feature subsets.

\begin{figure}[t]
    \includegraphics[width=.95\linewidth]{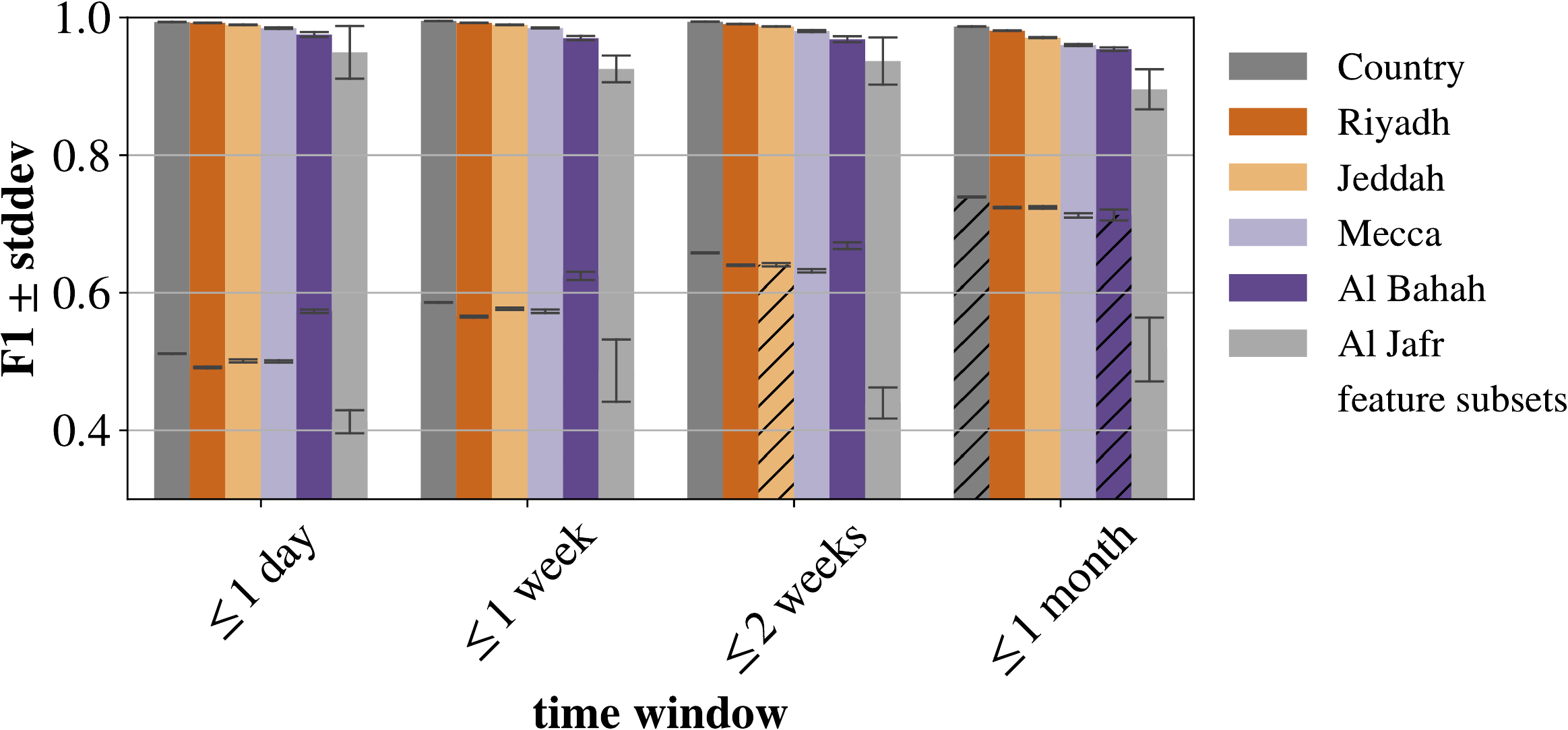}
    \caption{RF Binary Classification performance results.
        Regardless of observation time, the binary lifetime prediction works exceedingly well.
        All time-dependent feature subset comparisons perform by far worse.}
    \label{fig:bin_classifaction}
\end{figure}

We show the results of the best binary classification models in Figure~\ref{fig:bin_classifaction}.
The x-axis describes the used time window, whereas the y-axsis denotes the F1 score. %
The model results for each indepdendent community are plotted as bars with whiskers indicating the standard deviation across folds.
For comparison, we added feature subset analysis results (hatched).

We first observe that the binary classification model works quite well with F1 scores above $0.95$ for almost all communities across the time window.
The performance delta to the cumulative time-dependent feature subset analysis models only becomes smaller for longer time frames as those models improve.
With regards to model complexity, we observe similar results as seen for the other models (cf.~\ref{sec:model_sweetspot}): That is, \eg{} a tree depth of 16 performs far better than 8, but there are diminishing returns beyond this depth and more than 16 estimators.
In summary, the binary practical classifier outperforms \emph{any} other presented classification model.

\takeaway{In practice, it is desired to remove complete time-dependency.
Thus, the overall models may not reflect real-world performance in windowed feature use-cases.
We however showed that classifications on time-dependent window subsets do not perform well.
To ease up this problem, we simplified the task to a binary classifier predicting a users' lifetime.
This approach achieves better prediction quality than any other presented classifier.
}

\section{Related Work}
\label{sec:RW}
User churn prediction has been a research topic for decades, yet with new emerging use-cases and technical advancement in data mining, machine learning and explainable AI, research on this topic has not halted by any means.
We have seen various settings and applications within, \eg{} 
telecommnication~\cite{Oskarsdottir2020},
social networks~\cite{Chen2015},
online/video gaming~\cite{Runge2014,Chen2018},
or online marketing~\cite{Chamberlain2017}.
User lifetime has also been modelled as a survival analysis~\cite{Perianez2017}.

While user churn prediction often describes a binary classificaton, users' retention time might also be of interest.
From a marketing perspective, user churn measures are typically weighted into an optimization target of a Customer Lifetime Value (CLV) according to, \eg{} profit, yet relying on the same building block \cite{Fader2005, Chamberlain2017}.

Early work focussed less on today's off-the-shelv ML techniques.
That is, statistical modelling and distribution fitting has shown significant success~\cite{Fader2005a,Dror2012,Chen2015}.
Besides applying Markov models~\cite{Runge2014}, others have evaluated, \eg{} evolutional~\cite{WaiHoAu2003} or relational~\cite{Fader2005a,Oskarsdottir2020} minging techniques.
Nonetheless, various classical ML approaches have shown promising to very strong results with, \eg{}
boosting~\cite{Lemmens2006},
DTs \& tree ensembles~\cite{Dror2012,DanescuNiculescuMizil2013,Pudipeddi2014}%
.
Neural networks have also been applied to the problem in various architectures: \Eg{} deep~\cite{Yang2018,Chamberlain2017} or convolutional~\cite{Chen2018} NNs.
Explicit feature engineering for the data-driven methods requires an individual process to the very field of application.
However, research suggests that social ties and graphs are an important information carrier~\cite{Dror2012,Yang2018}.
Some research adds specific building blocks into their ML pipeline, such as user embeddings from browsing sessions~\cite{Chamberlain2017}.
Although, \eg{} RF or DT importances are often given, there is lack of its discussion, \ie{} backfed empirical implications are seldomly drawn.
Explainable AI is currently more or less tackled by, \eg{} applying a user pre-clustering~\cite{Yang2018}, however.

\section{Conclusion}
In this paper, we analyze and predict the lifetime of a user in Jodel, a mobile only location-based messaging app.
Our results show that Random Forests models provide good prediction results for both, regression and classification tasks across a selection of individual communities of varying sizes throughout the Kingdom of Saudi Arabia.
When making models invariant to total observation time, \ie{} only relying on timely limited feature sets, prediction results deteriorate substantially.
This can be solved by using Random Forrest models to predict a simpler binary classifcation problem of practical relevance to network operators: Given an observation time period, will the users’ lifetime be longer? 
This approach achieves even better prediction quality than any other presented classifier.

The location-based nature of Jodel yields the creation of disjoint communities throughout the country.
When training a single model to the entire data set (i.e., a country-wide model), this model performs well compared to individual community models at similar amounts of input data.
That is, while the individual communties are disjoint, users share behavioral pattern.
This is further highlighted by the fact that the RF feature importances correlate between most individual and the country model(s).
We therefore conclude the models' internal decision-making processes being similar and hence, also communities sharing alike behavior \wrt{} user liftime.

Eventually, we argue that the feature importance provides strong hints about model internals and are a good starting point to be fed back into empirical analyses, which we leave for future work.

\balance

\bibliographystyle{ACM-Reference-Format}
\bibliography{references}

\end{document}